\documentclass{article}

\usepackage[utf8]{inputenc} % allow utf-8 input
\usepackage[T1]{fontenc}    % use 8-bit T1 fonts
\usepackage{hyperref}       % hyperlinks
\usepackage{url}            % simple URL typesetting
\usepackage{booktabs}       % professional-quality tables
\usepackage{amsfonts}       % blackboard math symbols
\usepackage{nicefrac}       % compact symbols for 1/2, etc.
\usepackage{microtype}      % microtypography
\usepackage{xcolor}         % colors
\usepackage{graphicx}
\usepackage{amsmath}
\usepackage{wrapfig}
\PassOptionsToPackage{options}{natbib}
\usepackage[nonatbib]{neurips_2024}
\usepackage{multicol}
\usepackage{enumitem}
\usepackage{array}
\usepackage{threeparttable}
\usepackage{rotating}
\usepackage{caption}
\usepackage{makecell}
\usepackage{lineno}
\newcommand{\PreserveBackslash}[1]{\let\temp=\\#1\let\\=\temp}
\newcolumntype{C}[1]{>{\PreserveBackslash\centering}p{#1}}
\newcolumntype{R}[1]{>{\PreserveBackslash\raggedleft}p{#1}}
\newcolumntype{L}[1]{>{\PreserveBackslash\raggedright}p{#1}}

\title{Takin: A Cohort of Superior Quality Zero-shot Speech Generation Models}

\author{%
  David S.~Hippocampus\thanks{Use footnote for providing further information
    about author (webpage, alternative address)---\emph{not} for acknowledging
    funding agencies.} \\
  Department of Computer Science\\
  Cranberry-Lemon University\\
  Pittsburgh, PA 15213 \\
  \texttt{hippo@cs.cranberry-lemon.edu} \\
}

\begin{document}

\maketitle

\begin{abstract}
With the advent of the big data and large language model era, zero-shot personalized rapid customization has emerged as a significant trend.  
In this report, we introduce Takin AudioLLM, a series of techniques and models, mainly including Takin TTS, Takin VC, and Takin Morphing, specifically designed for audiobook production.
These models are capable of zero-shot speech production, generating high-quality speech that is nearly indistinguishable from real human speech and facilitating individuals to customize the speech content according to their own needs. 
Specifically, we first introduce Takin TTS, a neural codec language model that builds upon an enhanced neural speech codec and a multi-task training framework, capable of generating high-fidelity natural speech in a zero-shot way.
For Takin VC, we advocate an effective content and timbre joint modeling approach to improve the speaker similarity, while advocating for a conditional flow matching based decoder to further enhance its naturalness and expressiveness. 
Last, we propose the Takin Morphing system with highly decoupled and advanced timbre and prosody modeling approaches, which enables individuals to customize speech production with their preferred timbre and prosody in a precise and controllable manner.
Extensive experiments validate the effectiveness and robustness of our Takin AudioLLM series models. For detailed demos, please refer to  \url{https://everest-ai.github.io/takinaudiollm/}.

\end{abstract}

\section{Introduction}

Recent advancements in large language models (LLMs) \cite{brown2020language,openai2024gpt4,chu2024qwen2audiotechnicalreport,glm2024chatglmfamilylargelanguage}, neural codecs \cite{garbacea2019low,zeghidour2021soundstream,defossez2022high,pan2024promptcodechighfidelityneuralspeech}, and diffusion and flow models \cite{ho2020denoising,song2020denoising,rombach2022high,lipman2022flow,tong2023conditional} have led to significant progress in the fields of zero-shot text-to-speech synthesis (TTS) \cite{wang2023neural,lajszczak2024base,betker2023better,naturalspeech,naturalspeech2}, voice conversion (VC) \cite{zhao2022disentangling,dang2022training,wang2023lm,yao2024promptvc}, and related areas.  These innovations enable the synthesis of high-quality speech without extensive model training, thereby enhancing the accessibility and scalability of these technologies and fostering more natural and immersive user interactions.

In this context, to drive innovation and support audiobook production, we propose Takin AudioLLM—a series of models designed to allow users to customize speech content according to their specific needs while generating high-quality, near-human-like speech with exceptional naturalness and expressiveness. The Takin AudioLLM series comprises Takin TTS, Takin VC and Takin Morphing.

Firstly, inspired by the powerful contextual learning capabilities of LLMs, we present Takin TTS—a robust and effective neural codec language model for audiobook production. Takin TTS incorporates a high-fidelity, low-bandwidth neural speech codec based on efficient disentangled prompt encoders, which reduces modality heterogeneity between text and audio, thereby enhancing the LM's prediction accuracy. We introduce a five-stage multi-task training strategy that significantly improves overall LM performance, ensuring robustness and effectiveness in complex real-world scenarios. Additionally, we employ a latent diffusion model and Vocoder for token-to-speech synthesis, further improving speech quality and naturalness. Consequently, Takin TTS excels in generating high-quality, natural-sounding speech for various applications, from interactive voice response systems to sophisticated text-to-speech frameworks. This approach greatly enhances user experience and demonstrates substantial potential in advancing generative speech modeling technology.

Secondly, Takin-VC employs a joint modeling approach that integrates timbre features with both supervised and self-supervised content representations to enhance speaker similarity and intelligibility. This design allows Takin-VC to effectively capture and reproduce the nuanced characteristics of various speakers, ensuring that converted voices closely resemble the target speakers. Furthermore, to refine speech quality and naturalness, we incorporate an efficient conditional flow matching-based decoder. This advanced decoder optimizes the alignment between timbre and content features, leading to more accurate and natural voice conversion. In this way, Takin-VC provides a powerful and versatile tool for voice conversion applications, excelling in producing high-fidelity, natural-sounding voice conversions suitable for audiobook production. It significantly enhances user experience and demonstrates its potential to advance the field of voice conversion technology.

Finally, Takin Morphing introduces an attention mechanism-based multi-reference timbre encoder for precise and detailed timbre modeling. Additionally, a language model (LM)-based prosody encoder is employed to capture prosody representations that align with timbres for unseen speakers in an auto-regressive manner. To further enhance waveform quality, we advocate a two-stage information-flow-based training method. Through these innovations, Takin Morphing enables users to utilize timbres from various unseen speakers and combine them with preferred prosody styles, thus generating personalized audiobooks with a high degree of control. This capability meets the demands of diverse speech synthesis applications, from entertainment and education to commercial contexts, offering a more natural and enriched auditory experience.

Overall, Takin AudioLLM represents a significant advancement in zero-shot speech production technology. By leveraging the sophisticated capabilities of Takin TTS, Takin VC, and Takin Morphing, this series not only advances the state-of-the-art in speech synthesis but also addresses the growing demand for personalized audiobook production, enabling users to tailor speech generation precisely to their requirements.

% \section{Related Work}

\section{Takin TTS}

\subsection{Overview}

\begin{figure}[h]
    \centering
    \includegraphics[width=14cm]{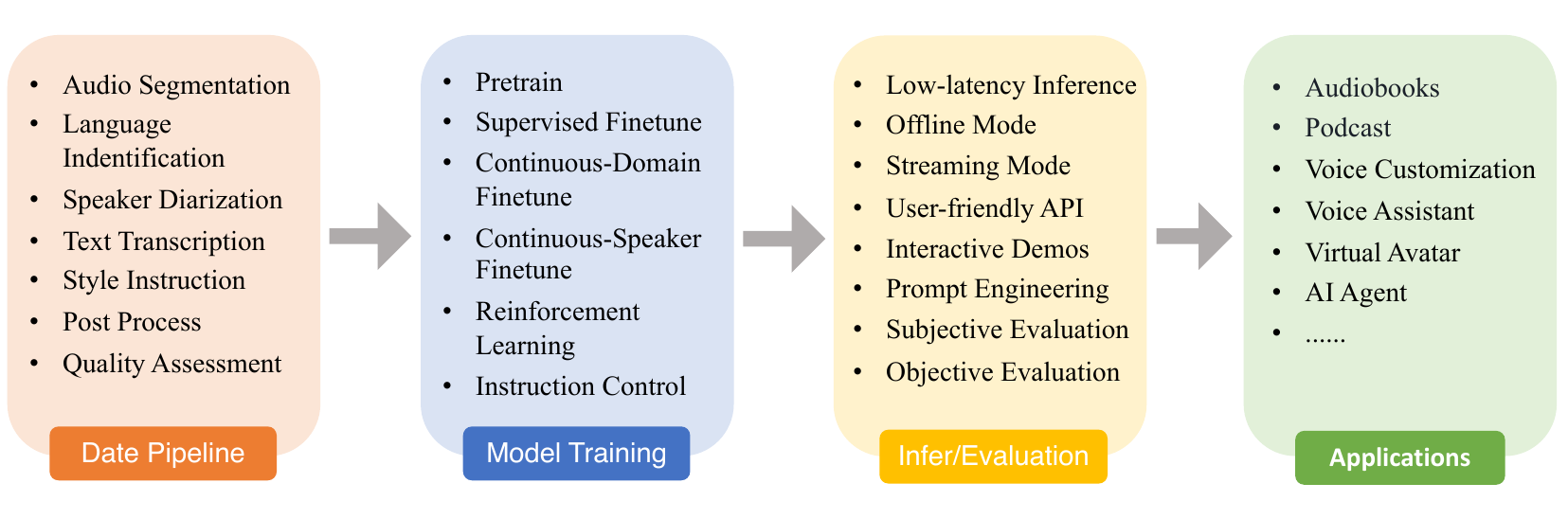}
    \caption{An overview of Takin TTS workflow}
    \label{fig:overview}
\end{figure}

\begin{figure}[h]
    \centering
    \includegraphics[width=14cm]{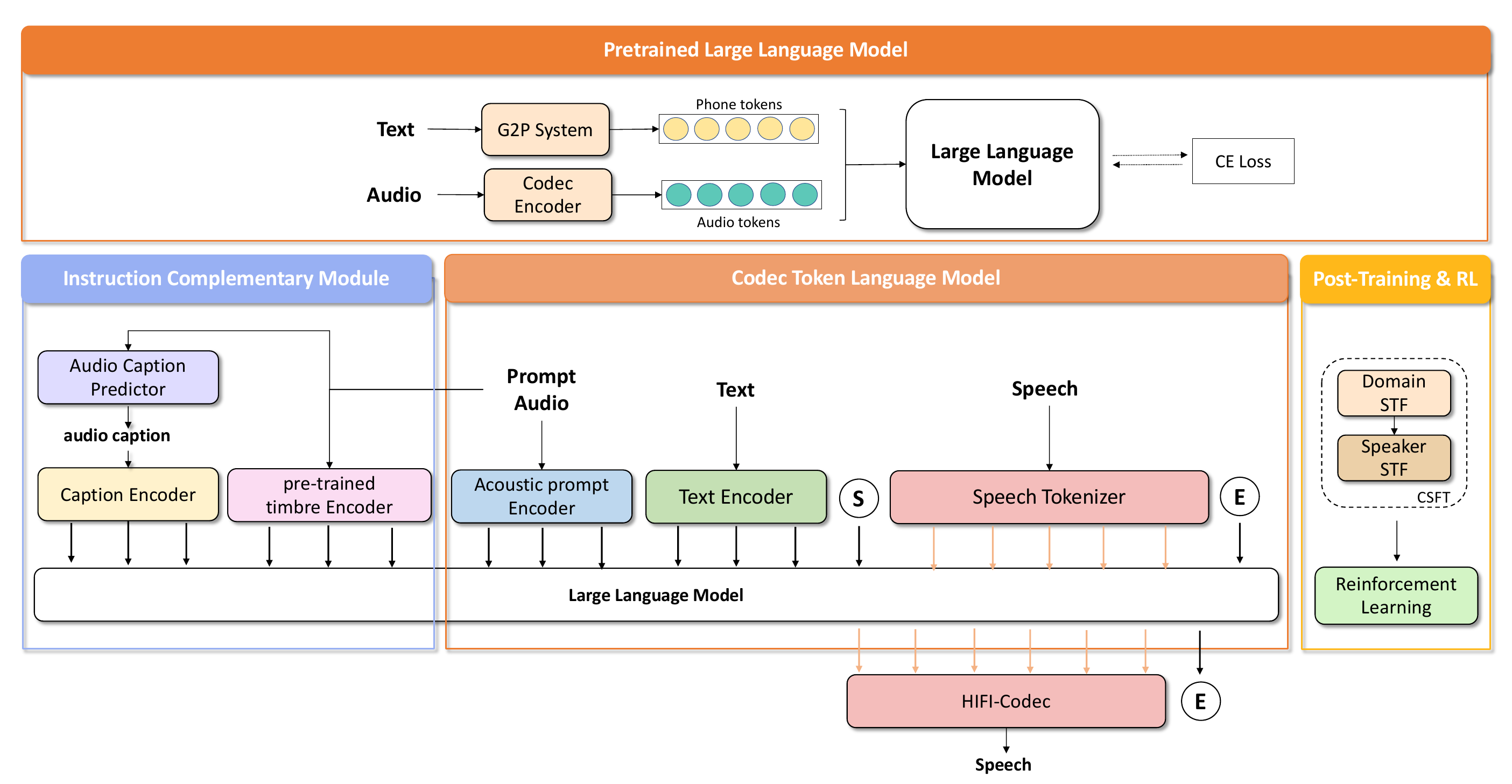}
    \caption{Overall schematic diagram of Takin TTS.}
    \label{fig:takin-tts}
\end{figure}

As shown in Figure \ref{fig:overview}, we take Takin TTS as an example to introduce the construction scheme of this series of large models, primarily including the construction of large-scale datasets, model training for specific tasks, the establishment of evaluation systems for voice generation models, and the commercialization of applications.
Additionally, Figure \ref{fig:takin-tts} illustrates the overall training process of Takin TTS, and the specific training details will be gradually expanded upon as follows.
% The overall process of Takin TTS training is shown in Figure \ref{fig:takin-tts}, and the specific training details will be gradually expanded upon in the following chapters.

\subsection{Pretrain}
% We use multimodal information to pre-train the model. Specifically, we encode the text and audio data pairs into tokens and output them to the GPT model to learn the knowledge.

We use multimodal data to pretrain the Takin TTS. Specifically, we encode text and audio data into tokens and input them into the GPT model to learn relevant knowledge. For text data, we develop an internally developed G2P (Grapheme-to-Phoneme) method. This solution includes a Text Normalization (TN) module, a Named Entity Recognition (NER) module, as well as a polyphone disambiguation module, which can convert text into phonemes and subsequently embed them into the lexical merge space. For audio data, we train an encodec system with a single codebook to convert audio content into discrete codec tokens. In this way, the multimodal data can be input into the GPT model for its understanding.

\begin{equation}
\label{eq:ntp}
    \hat{x}_t = \arg\max_{x_t} P_\theta(x_t \mid x_1, x_2, ..., x_{t-1})
\end{equation}

We use the most classic GPT training method, assuming the audio or text sequence is $X=\{x_1, x_2, ..., x_t\}$, with the pretraining objective as shown in equation \ref{eq:ntp}.

% \begin{figure}[h]
%     \centering
%     \includegraphics[width=10cm]{figures/pretrain.pdf}
%     \caption{Pretrain Process of Takin Audio Large Language Model}
%     \label{fig:enter-label}
% \end{figure}

\subsection{Supervised Fine-tuning (SFT)}

Following unsupervised learning on extensive data, our Takin TTS has developed a robust capacity to comprehend text and audio information.
In the subsequent phase, akin to GPT-4 \cite{openai2024gpt4}, we employ labeled paired data to train the Takin TTS model for downstream tasks such as TTS and Automatic Speech Recognition (ASR) \cite{gulati2020conformer,kim2022squeezeformer,yang2022lmec}, thereby enhancing its proficiency in managing text and speech tasks.

In the TTS task, zero-shot is a quite important capability of applications that requires the model to synthesize high-quality speech for unseen speakers without collecting their labeled data for training in advance. In this work, leveraging the ability of neural codec to convert speech into discrete tokens, the zero-shot TTS task is regarded as a conditional language modeling task to predict discrete codec tokens autoregressively based on given conditions.

Let $D=\{T_i, P_i, S_i\}$ denotes the training dataset, where $S_i$ is the target speech, $T_i$ is the text description and $P_i$ is prompt audio which is from the same speaker with $S_i$. During the training process, a set of speech conditions $SC_i$ is extracted from prompt audio $P_i$ via acoustic prompt encoder, the text transcription $T_i$ is converted to a phoneme sequence $TP_i = \{BP, P_{i_1}, P_{i_2}, ...,P_{i_m}, EP \}$, and $BP$ stands for the Begin of Phone Sequence, $EP$ stands for the End of Phone, the target speech is passed to neural codec model to get discrete codec tokens $C_i = \{ C_{i_1}, C_{i_2}, ...,C_{i_n} \}$. A start identifier $s$ and an end identifier $e$ are inserted at the beginning and the end of codec tokens. The input training sequence is constructed as follows:
$$[SC_i, TP_i, S, C_i, E]$$

As shown in Figure \ref{fig:takin-tts}, the language model is only trained to predict codec tokens and the end of sequence token E conditioned on phone sequence $TP_i$ and speech conditions $SC_i$, which is formulated as:
\begin{equation}
  P(C_i |S, SC_i, TP_i )= \prod \limits_{t=1}^{n+1} P(C_{i_t} |C_{i_<t}, S, SC_i, TP_i)  
\end{equation}
% $$P(C_i |S, SC_i, TP_i )= \prod \limits_{t=1}^{n+1} P(C_{i_t} |C_{i_<t}, S, SC_i, TP_i)$$
where $C_{n+1}$ denotes the end of sequence token E. During inference, the language model generates tokens autoregressively based on given text and reference speech, and fed these tokens to neural codec model to generate audio.

% The input of the SFT Process is [$\{<SC_1>, <SC_2>, ..., <SC_{N_C}> | <BP>, <P1>, <P2>, ...,<P_{N_P}>, <EP>|<BC>, <C1>, <C2>, ..., <C_{N_C}>, <EC>\}$, where $S_C$ stands for Speech Conditions, $BP$ stands for Begin of Phone Sequence, $EP$ stands for End of Phone. Similar to the Phone sequence, $BC$ stands for Begin of Codec token, and $EC$ stands for the End.

\subsection{Continual Supervised Fine-tuning (CSFT)} \label{method:csft}
While Supervised Fine-Tuning (SFT) has endowed the Takin TTS model with TTS capabilities, the diverse content standards generated by TTS often lead to more frequent word omissions in Autoregressive (AR) models compared to Non-Autoregressive (NAR) models during inference \cite{ren2022revisiting,borsos2023soundstorm}. As a consequence, to enhance the stability of the system's TTS functionality, further Continual Supervised Fine-tuning (CSFT) joint with ASR guided training is necessary.
In our method, CSFT primarily consists of two components: Domain-SFT and Speaker-SFT, which will be elaborated below.

% TODO: 加一个小图，体现强化学习和ASR Joint Training @jingjing @ yuguang

% \begin{figure}[h]
%     \centering
%     \includegraphics[width=12cm]{figures/CSFT.png}
%     \caption{Training Process of Continual Supervised Fine-tuning}
%     \label{fig:enter-label}
% \end{figure}

\subsubsection{Domain SFT}

% 方法介绍：
% - MOE - LoRA stack 1
% 测试结果：
% - 证明style更加贴近domain

% Our extensive experiments demonstrate that Domain SFT can further improve the quality and accuracy of generated speech. 

% LoRAMoE参考：https://arxiv.org/pdf/2312.09979

\begin{figure}[h]
    \centering
    \includegraphics[width=10cm]{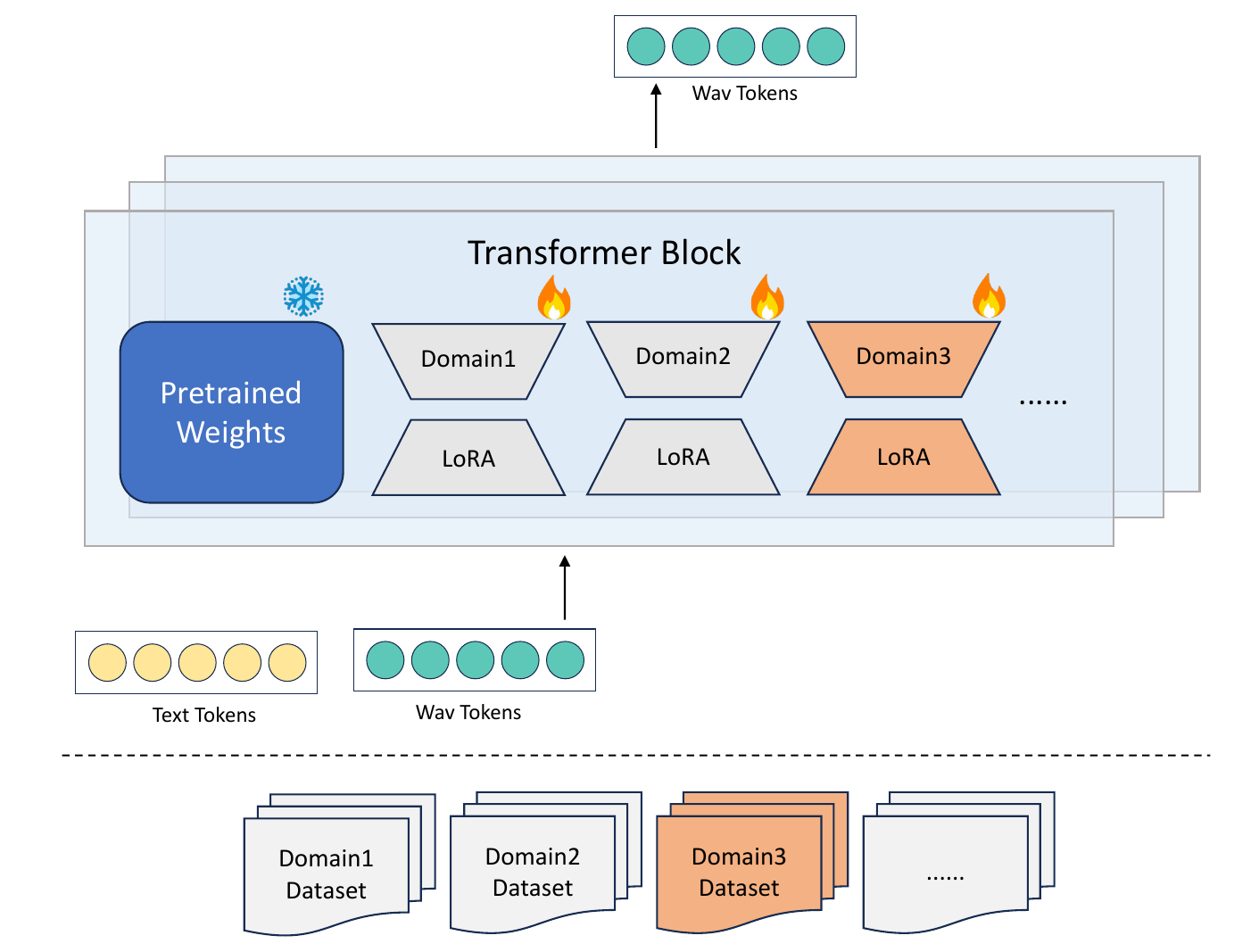}
    \caption{LoRAMoE for Domain Fine-tuning}
    \label{fig:lora-moe}
\end{figure}

The distribution of speech prosody diverses across various scenarios, for instance, the overall prosody of reading an audiobook is far different from that of delivering a speech. Since there usually exists an imbalanced data during pretraining phase, in order to improve the quality and accuracy of generated speech, Domain SFT is applied to our well fine-tuned models. In this phase, we only select several thousand hours of high-quality finely labeled domain data and train the propsoed approach using LoRA \cite{hu2021lora}.

\subsubsection{Speaker SFT}

% 方法介绍：
% - 基于Domain SFT做延伸描述，MOE - LoRA stack2；可实现 Domain SFT + Speaker SFT
% 测试结果：
% - Domain SFT + Speaker SFT => 证明speaker similarity和intelligibility都有更好的提升

\begin{figure}[h]
    \centering
    \includegraphics[width=10cm]{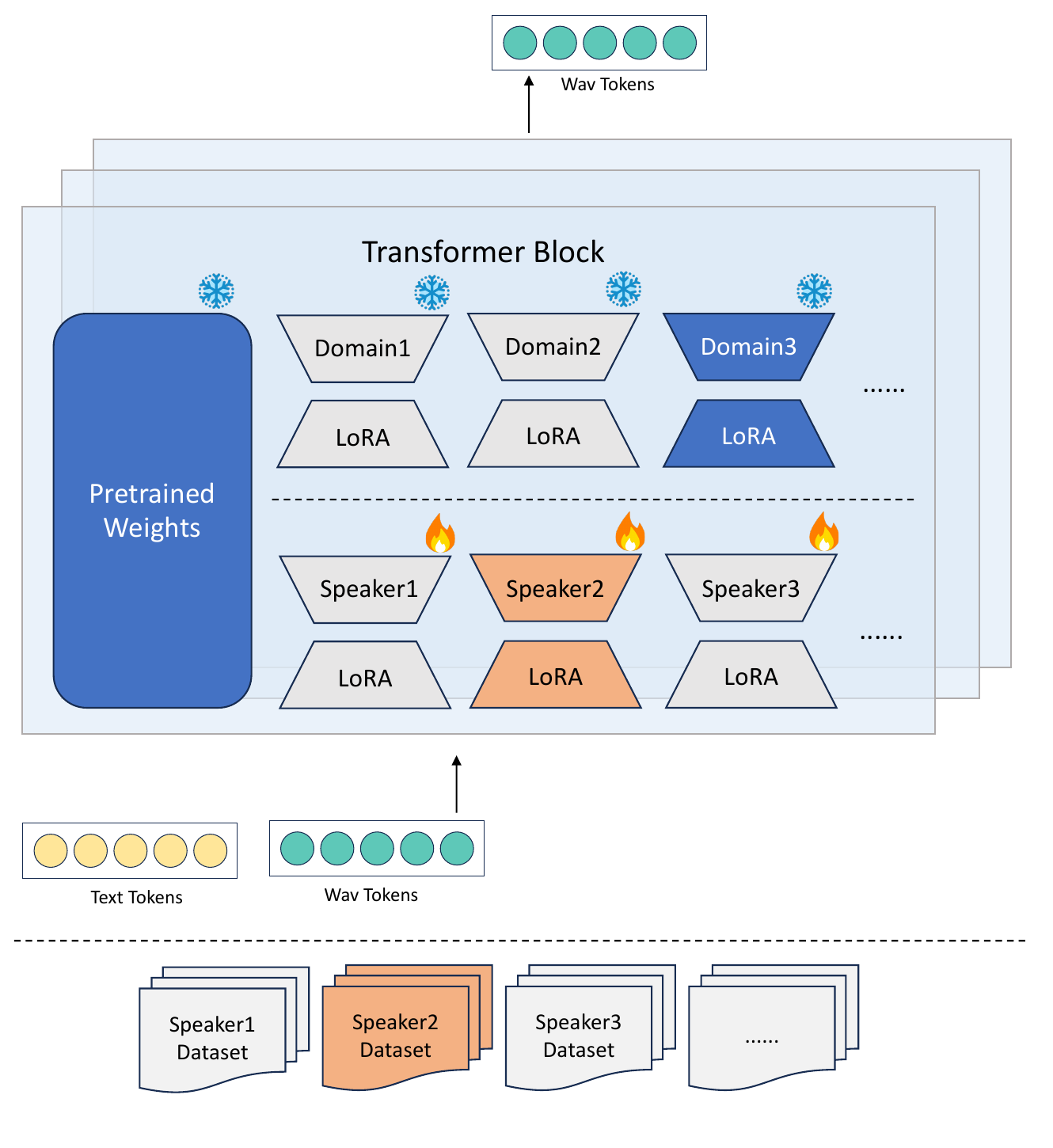}
    \caption{LoRAMoE for Speaker Fine-tuning}
    \label{fig:enter-label}
\end{figure}

% TODO: 内容扩充 @jingjing
To ensure that the narration of high-quality audiobooks sounds more natural and aligns closely with the original speaker's performance style, we have further introduced the Speaker SFT Phase. In this section, we continue to use the LoRA training method. The difference here is that we freeze most of the GPT parameters to retain the model's foundational knowledge and update the parameters of the Acoustic Prompt Encoder with the Input and Output Embedding Layer parts of GPT.

\subsubsection{ASR guided Joint Training}
To improve the accuracy of the model's output content, we incorporate ASR guidance into the model training during the finetuning process. The sequences output by GPT are fed into a codec decoder to be restored to wav format. To ensure gradient propagation and training speed, the generated wav is input into the whisper model, and its output is compared with the annotations to calculate the cross-entropy loss.

\subsubsection{Reinforcement Learning}

% 

% 方法介绍：
% - Domain SFT + Speaker SFT + RL
% 测试结果：
% - Domain SFT + Speaker SFT + RL => 证明有更低的bad case rate 和 expressiveness

Despite the fact that the model after CSFT Process performs quite well, even surpassing human rendition levels for certain sentences by some speakers, it still faces issues with varying effectiveness among different speakers for the same text, as well as discrepancies between human and machine aesthetics. To make the generated content as closely aligned with human preferences as possible, we have introduced the concept of our RL (Reinforcement Learning) method. As shown in Figure \ref{fig:takin-tts}, The RL is placed in the end of the whole diagram of Takin TTS to further improve the performance by aligning the model with human preference.

Currently RL methods \cite{dpo, schulman2017proximal, ethayarajh2024kto, jung2024binary},  follow a Sampling-human-annotating-learning pipeline, in which human evaluation is applied to model-generated outputs to ensure the model learns to align with subjective human preferences. The pipeline works also in speech generation task \cite{arxivSpeechAlignAligning}. However the human ratings is labour dmanded, there are also some works studying to \cite{arxivEnhancingZeroshot, arxivRobustZeroShot} use objective metrics to replace human ratings, in order to facilitate the obtaining of preference data pairs. We also explore leveraging the human-rating only pipeline to combine it with a set of objective metrics which partly indicate human preferences, namely the Sampling-human\&machine-annotating-learning pipeline.

% The RL is placed in the end of the whole diagram of Takin TTS in Figure {fig:takin-tts}  By incorporating this pipeline to the model, it  offers a viable strategy to
% address the discrepancy in history sequence between TTS training and test, as it exposes the model with the samples from self-generated history during training and it further leads to improved TTS robustness and gains in expressiveness.

% TODO: 图片修改 @jingjing
% \begin{figure}[h]
%     \centering
%     \includegraphics[width=10cm]{figures/pretrain.pdf}
%     \caption{The Pipeline of Reinforcement Learning}
%     \label{fig:rl-1}
% \end{figure}

\subsubsection{Instruction Style Control}
% TODO: 图片补充、内容扩充到SFT篇幅左右、补充公式？

In AudioLLM paradigm, the common method of controlling speech style involves selecting different audio prompts, which generally incorporate both speaker identity and style information simultaneously. To explore the full potential of controllability, we propose TakinTTS-Instruct to synthesize speech with various styles and emotions, including rhythm, pitch, paralinguistics, etc., using natural language as style prompts which is more user-friendly than the base model of Takin TTS and could decouple the speaker and style in synthesis.

The lower-left part of Figure \ref{fig:takin-tts} prominently displays the core structure of TakinTTS-Instruct.
To be specific, a robust pre-trained speaker verification system \cite{wang2023cam++} is employed to provide additional voice characteristics, enhancing the similarity between the synthesized voice and the target speaker. Moreover, unlike previous speech emotion or speaking state recognition tasks \cite{10448,2024023,2024012,2023yi}, in order to control the emotions of the generated speeches, speaking states, or other linguistic dimensions in a more user-friendly fashion, we implement a predictor that detects and classifies emotions or different speaker states in spoken language and subsequently outputs its corresponding natural language description. Furthermore, these descriptions will be parsed by SimBERT\cite{simbert} into an embedding form to be incorporated into the model training.

\section{Takin VC}
In addition to TTS, another widely used technology in the audiobook business is VC technology. Here, we propose a novel and effective zero-shot VC approach based on DDPM or CFM. Similar to the usage conditions of TTS technology, it can achieve high-expressiveness timbre conversion with only 5-10 seconds of unseen audio.

% The structure of Takin VC is as follows: one part of its input is Phonetic Posteriorgrams PPG (in this case, we are using the output from HybridFormer), and the other part is a truncated prompt mel fragment. As for the output, we have two options: one is to output as a mel spectrogram and then restore it to a wav through a vocoder, and the other is to directly restore it to a wav.

\subsection{VC Training}

The input of Takin VC is composed of two parts: Phonetic Posteriorgrams (PPG), utilizing the output features of HybridFormer \cite{yang2023hybridformer} in this case, and a truncated prompt mel fragment. For the output, Takin VC offers two alternatives: it can either produce mel spectrograms, which are subsequently converted to audio samples through a vocoder, or directly generate audio samples. 

\begin{figure}[h]
    \centering
    \includegraphics[width=14cm]{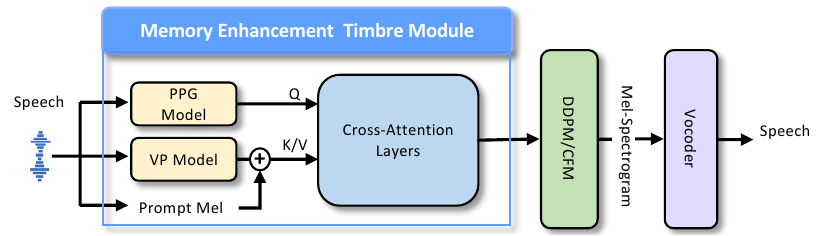}
    \caption{The Structure of Takin VC.}
    \label{fig:takin-vc}
\end{figure}

% TODO: 细化介绍上图VC框架的流程
To elaborate, our Takin VC system mainly consists of three components: the PPG model
is used to extract the decoupled content information of the input audio, and the memory augmented cross-attention based timbre modeling mechanism is used to re-populate the timbre information. Finally, we use the way of DDPM / Flow Matching to restore the converted spectral information and employ a HiFi-Gan vocoder to render it into the converted audio. Specifically, the PPG Model used here is a pre-trained Hybridformer model. As for the voiceprint model, we use a pre-trained CAM++\cite{wang2023cam++} model from modelscope. The target features of DDPM / CFM here use a 129-dimensional mel-filterbank, and the prompt mel does the same.  Due to the difficulty of collecting VC data from real world, similar to other VC system, we directly use normal speech data by cut of single speaker, and randomly select a segment as the prompt mel. We trained our models on 500k hours of data. During training , the PPG model and voiceprint model are frozen and only update the memory augmented timbre blocks and DDPM / CFM model are updated.

% \subsubsection{Discrete Timestep Training}

\subsection{VC CSFT}

Similar to parts of the Takin TTS, after pre-training on a large amount of data, fine-tuning with a small amount of high-quality data can enhance the model's performance. Furthermore, although the timbre information retained in the PPG feature is already minimal, there are still minor timbre leakage issues, resulting in suboptimal timbre conversion similarity in some cases. Therefore, we employed the TTS system to improve the performance in this regard. Due to the duration control issues, we used a traditional TTS system here to generate a small amount of parallel data, which is used to better guide the model in understanding the speech conversion task.

\section{Takin Morphing}

% Style Transfer is an important application in the field of audiobook production. It involves converting styles while preserving the speaker's tone, thereby lowering the threshold for becoming a professional narrator. Using this technology, works by enthusiasts who are not yet skilled in broadcasting techniques can be transformed into the style of professional narrators, enhancing the quality of the pieces to a certain extent. It can also serve as a teaching aid to guide enthusiasts in developing their own unique broadcasting styles.

Audio Style Transfer is an important application in the field of audiobook production, which involves transforming styles while retaining the speaker's vocal timbre, thereby lowering the barrier to becoming a professional broadcaster. By using this technology, works by enthusiasts who are not yet proficient in certain broadcasting techniques can be transformed to have the style of professional broadcasters, thereby improving the quality of the works to some extent. Alternatively, it can serve as an auxiliary in teaching, guiding enthusiasts to develop their own unique broadcasting styles. Here, we introduce the Takin Morphing technology, which utilizes the form of DDPM to achieve style and rhythm transfer while maintaining the speaker's vocal timbre.

\begin{figure}[h]
    \centering
    \includegraphics[width=14cm]{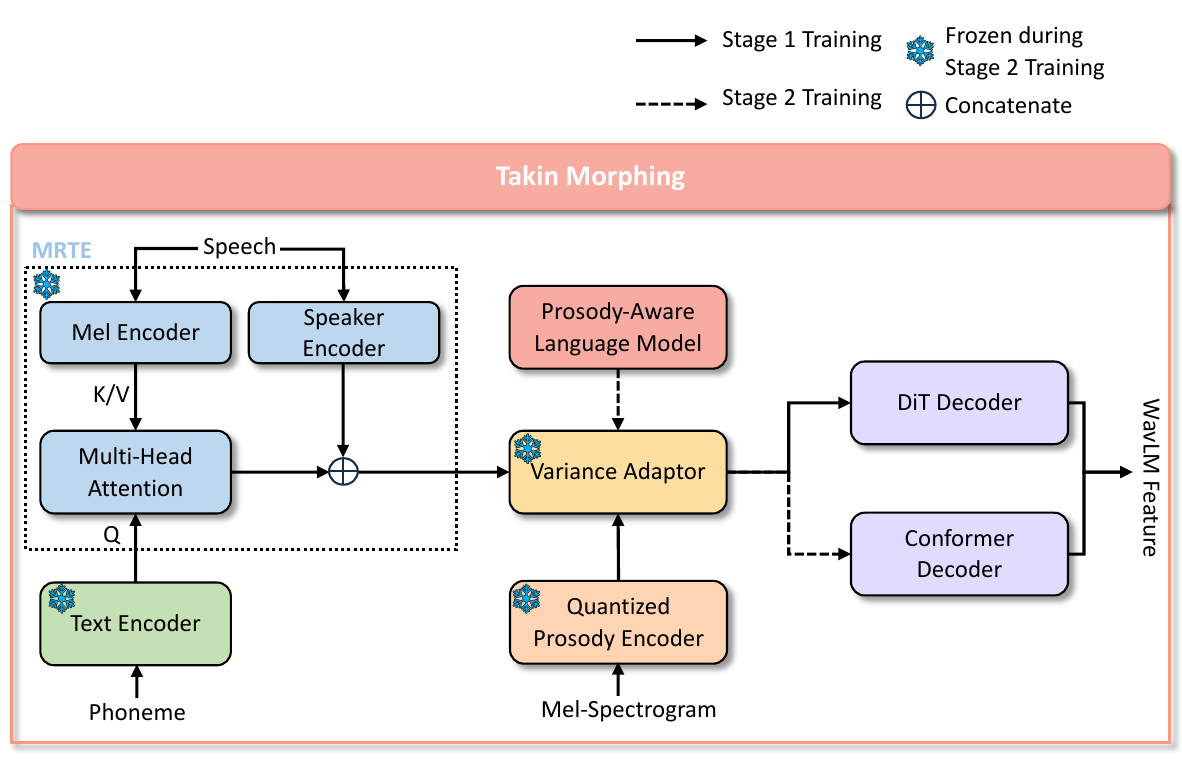}
    \caption{Training Structure of Takin Morphing.}
    \label{fig:audio-morphing}
\end{figure}

As shown in the Figure \ref{fig:audio-morphing}, the input to the model is the Phone sequence. After feeding it into the Multi-Reference Timbre Encoder (MRTE) layer \cite{megatts}, we obtain a hidden matrix containing content and timbre information. This matrix, along with prosodic features, is sent into the Decoder model to restore the mel-like features. The prosodic features are VQ vectors of low-frequency Mel. Here, we replaced the commonly used mel-spectral features with wavlm features because our experiments show that wavlm features contains more expressive information than Mel Spectrum.

\section{Experiments}

\subsection{Takin TTS Settings}

\par\noindent\textbf{Experimental Datasets} To train and evaluate Takin TTS, We build a large multilingual base dataset for pretraining and 1st round of SFT training. To evaluate CSFT and RL training, several carefully human labeled datasets are built which including domain datasets and speaker datasets. The datasets are depicted as follows: 

\begin{itemize} 
  \item \textbf{Base TTS Dataset}: In-house dataset including over 1M hours of speech data, which may include some labelling errors.
  %\item \textbf{high-quality} Dataset: 2000 hours of high-quality speech data, with all the transcripts manually checked and the amount of data for each speaker is at least over 1 hour.  
  \item \textbf{Domain TTS Dataset}: The domain dataset is of high-quality dataset with all the transcripts manually checked. There are two domains exist in the domain dataset which are audiobook and podcast. There is around 1000 hours for each domain used for Domain SFT. For speech data of each domain, 5\% of the whole dataset is held out for test purpose and we make sure the held out test set does not have speaker overlap with the train set.
  \item \textbf{Speaker TTS Dataset}: To conduct speaker SFT based on the result of Domain SFT, a small Speaker TTS Dataset is constructed by selecting two audiobook-domain speakers and two podcast-domain speakers from our in-house dastaset. There is 1-hour speech data for each speaker, and likewise their transcripts are carefully labeled. For each speaker, 5\% of speech data is held out as test set.
  % TODO: add domain dataset description
\end{itemize}

\par\noindent\textbf{Evaluation Metrics} To conduct objective evaluations, We employ the Phoneme Error Rate (PER) and Speaker Similarity (SIM) metrics. For PER, we pick Whisper-large-v3 \cite{whisper} as the ASR model to conduct the PER test, While for SIM, we use CAM++ on the speaker verification task \cite{camplusplus} to obtain speaker embeddings for calculating the cosine similarity of speech samples of each test utterance against reference clips. For subjective evaluations, We employ the Mean Opinion Scores (MOS) by rating different speech samples of the same content by human evaluators. The scores vary from 1 to 5 and the higher score indicates better speech quality. Besides, Bad Case Rate (BCR) is used to evaluate the overall stability of our models in RL experiments. Equation \ref{eq:bcr} is defined to compute BCR, in which $B$ is the number of bad cases. To count the number of bad cases, We count the occurances of three types of bad cases covering prosody, pronunciation and missing or extra speech.

\begin{equation}
  BCR = \frac{B}{100}
  \label{eq:bcr}
\end{equation}

% \subsubsection{Takin TTS Datasets}
% \subsubsection{Multi-modal Pretraining}

% \subsubsection{Takin TTS Pretraining}

\subsubsection{Pretraining}

% 可以统计一下PPL？

\begin{figure}[h]
    \centering
    \includegraphics[width=14cm]{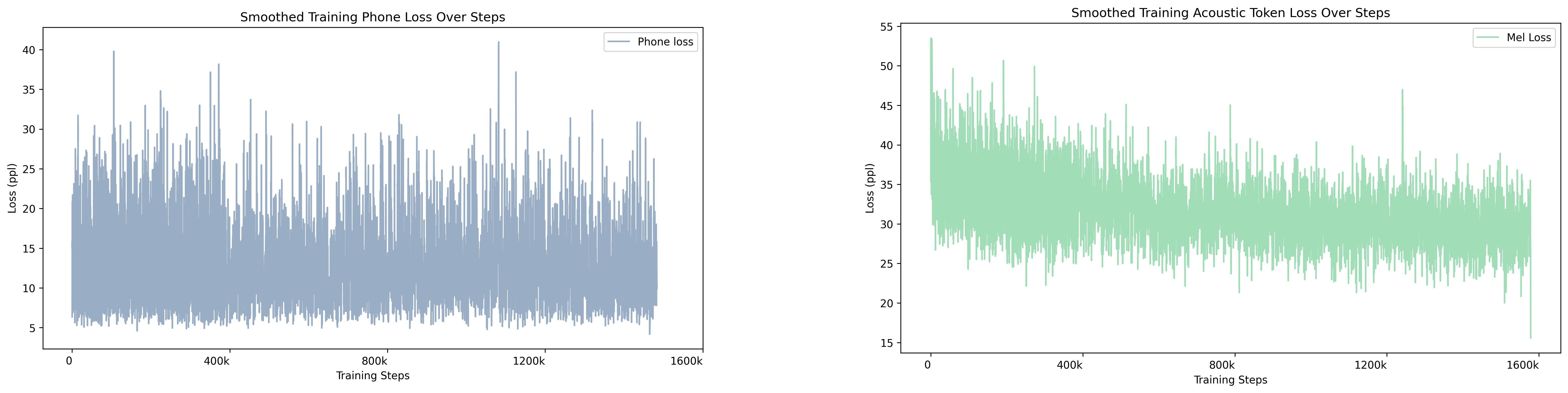}
    \caption{Training Loss of Phone and Acoustic Tokens.}
    \label{fig:pretrain_loss}
\end{figure}

As shown in the Figure \ref{fig:pretrain_loss}, during the pretraining process, the phone loss can converge quickly. In comparison, the learning of acoustic tokens is slower, seemingly because the acoustic tokens contain more information and are more difficult to learn. Although we believe that if one only wants to perform TTS tasks, pre-training may not be a necessary option, and starting from the CSFT stage can also train a model with very good results. However, if you want the model to expand other multimodal capabilities, such as GPT-4o, pretraining is also a good choice.

\subsubsection{CSFT on Takin TTS}
% 参考 CosyVoice / Seed-TTS
% 客观指标：zero-shot能力：PER / SIM / DNSMOS
% 结构：参考seed-tts 粗写
% 追加英文测试结果

After pretraining, Talkin TTS model is finetuned with \textit{Base TTS dataset} to align the model to TTS task, namely SFT. However, this finetuned model is not prepared for real applications in terms of its stability and expressiveness as mentioned in section \ref{method:csft}. CSFT is key to getting a stable model and enhance its expressiveness, especially when generating speech of a specific speaker. This section is mainly focused on the experiments of two types of CSFT which are domain SFT and speaker SFT.

% \par\noindent\textbf{Experimental results of }: First, we utilize the high-quality dataset to fully finetune the whole takin TTS model with relatively small learning rate which is 1e-5. That SFT experiment dramatically helps the model to converge into a more stable status and provides a more natural expressiveness. The Table \ref{table:exp_csft} shows the PER of the model with SFT is superior to that without SFT, as well as the MOS score. For the zero-shot speaker similarity, both models with and without SFT share a similar SIM score. 

\begin{table}[h]
\caption{Evaluating results of domain SFT}
\centering
\begin{tabular}{cccc}
\hline
Model & PER($\downarrow$) & SIM($\uparrow$)  & MOS($\uparrow$)  \\ \hline
w/o domain SFT & 5.6 & 0.70 & 4.12 $\pm$ 0.09  \\
w/ domain SFT (cross-domain test) & 3.3 & 0.69 & 4.18 $\pm$ 0.07 \\ 
\textbf{w/ domain SFT (in-domain test)} & \textbf{2.8} & \textbf{0.71} & \textbf{4.29 $\pm$ 0.06} \\ \hline
\end{tabular}
\label{table:exp_domain_sft}
\end{table}

\par\noindent\textbf{Domain SFT}, unlike the full-parameter fine-tuning of CSFT, trains extra LoRA parameters of Takin TTS model, keeping backbone frozen. \textit{Domain Dataset} is used to do domain SFT on audiobook and podcast domains respectively, which consequently results in two separate LoRA models of those two domains. The experimental results are demonstrated in Table \ref{table:exp_domain_sft}, the model after SFT is denoted by \textit{w/o domain SFT}, and \textit{w/ domain SFT} denotes the SFT model further finetuned with domain SFT. All the objective metrics are computed in a zero-shot setup using test set in \textit{Domain TTS Dataset}. To conduct subjective evaluation, 8 males and 8 females are randomly selected from the test set to synthesize speech, with 30 subjects participated in to rate scores from 1 to 5. We not only compare the model with and without domain SFT, but also study the influence of domain match between training and inference. \textit{cross-domain test} labels testing a model after domain SFT with mismatched domain data. e.g., testing the model after audiobook domain SFT with podcast test data. On the contrary, \textit{in-domain} test denotes the testing scenario with matched domain data. The Table \ref{table:exp_domain_sft} shows the PER of the model with domain SFT is superior to that without domain SFT, as well as the MOS score. For the zero-shot speaker similarity, both models with and without domain SFT share a similar SIM score. We also find that the domain consistency of training and inference further boosts the performance of the generated speech from both subjective and objective perspectives.

% \subsubsection{CSFT Training on TTS}

\begin{table}[h]
\caption{Evaluating results of Speaker SFT}
\centering

\begin{tabular}{cccc}
\hline
                            Model    & PER($\downarrow$)                      & SIM($\uparrow$)                      & MOS($\uparrow$)                      \\ \hline
%domain SFT                            & 2.32 & 0.67 & 4.07 $\pm$ 0.08 \\ 
domain SFT               & 1.91                     & 0.70                     & 4.23 $\pm$ 0.06                    \\
speaker SFT               & 1.13                     & 0.81                     & 4.35 $\pm$ 0.08                    \\
\textbf{domain SFT + speaker SFT} & \textbf{0.89}                     & \textbf{0.82}                     & \textbf{4.46 $\pm$ 0.07}                    \\ \hline
\end{tabular}
\label{table:spk_sft}

\end{table}

\par\noindent\textbf{Speaker SFT} also trains extra LoRA parameters of Takin TTS model and can be stacked onto the model after Domain SFT. We conduct the experiment of speaker SFT based on \textit{Speaker TTS Dataset} and all the evaluations are based on the test data of the four speakers in that dataset. In the experiment of speaker SFT, we study the influence of stacking domain SFT and speaker SFT, by comparing it with the models only with domain SFT and the ones with only speaker SFT. 

As illustrated in Table \ref{table:spk_sft}, we use the same evaluations as that in domain SFT experiments. the speaker SFT obviously plays the most important part to improve the performance of a specific speaker on both objective and subjective perspectives. However, just speaker SFT alone does not achieve the best results. We suspect that is due to the model after domain SFT works like a better starting point for speaker SFT in the same domain.

\subsubsection{RL Training on TTS}

RL training can be employed as an extra post-training stage after either Domain SFT or Speaker SFT. Both experiments are conducted to verify the effectiveness of RL training, especially on expressiveness and BCR. To prepare training data for RL, we make a set of good / bad examples with both subjective ratings and objective metrics. As \cite{arxivLargeLanguage} shows repeated sampling is able to largely increase the pass coverage to queried problems, we get 5 samples by repeated sampling for each sentence. For objective ratings, we pick PER and UTMOS \cite{saeki2022utmos} as objective metrics to generate preferences considering both metrics. For subjective ratings, there are 50 human raters being participated in to rate the best and the worst sample among 5 candidates by listening and comparing the overall speech quality. Therefore, a good / bad example pair is acquired for each sentence either by objective or subjective ratings. As a result, for the RL experiments, 50000-sentence RL pairs are created for RL training after Domain SFT and 500-sentence RL pairs are prepared for two male and two female speakers respectively after speaker SFT. The experiments are defined as follows. 

\begin{itemize}
\item \par\noindent\textbf{\textit{Domain-SFT-RL-OBJ}} denotes the RL experiments conducted based on the model after domain SFT by using objective-metric-rating data.  
\item \par\noindent\textbf{\textit{Domain-Speaker-SFT-RL-OBJ}} denotes the RL experiments conducted based on the model after domain and speaker SFT by using objective-metric-rating data.  
\item \par\noindent\textbf{\textit{Domain-SFT-RL-SUBJ}} denotes the RL experiments conducted based on the model after domain SFT by using human-rating data. 
\item \par\noindent\textbf{\textit{Domain-Speaker-SFT-RL-SUBJ}} denotes the RL experiments conducted based on the model after speaker SFT by using human-rating data. 
\end{itemize}

% \par\noindent\textbf{DPO training}  

In our experiments, DPO is employed to do RL post training on \textit{Domain-SFT-RL} and \textit{Speaker-SFT-RL} respectively. The Table \ref{table:exp_rl} shows the results comparing models with (w/) or without (w/o) RL training. To make the result comparable, the four speakers in \textit{Speaker TTS Dataset} are picked to evaluate various metrics. The models named with \textit{SUBJ} suffix are trained with human-rating pairs, the results of which indicate more stable speech generation in terms of \textit{PER} and \textit{BCR}. However there is just minor improvements on MOS which is more related to expressiveness. That might be due to RL data raters more sensitive to bad cases comparing to prosody changes.

% Finding
% + uncertainty and consistency finding of human rating and objective metric rating

\begin{table}[h]
\caption{Objective and subject evaluating results of models with and without DPO.}
\centering
\begin{tabular}{lcccc}
\hline
Model & PER & BCR & MOS & SIM \\ 
\hline
Domain-SFT w/o RL & 1.91 & 1.1\% & 4.23 $\pm$ 0.06 & 0.69 \\
Domain-Speaker-SFT w/o RL & 0.89 & 0.7\% & 4.46 $\pm$ 0.07 & 0.82 \\ 
\hline
Domain-SFT-RL-SUBJ & 1.79 & 0.9\% & 4.26 $\pm$ 0.07 & 0.71 \\
Speaker-SFT-RL-SUBJ & 0.89 & 0.4\% & 4.53 $\pm$ 0.09 & 0.81 \\ 
\hline
Domain-SFT-RL-OBJ & 1.89 & 1.1\% & 4.22 $\pm$ 0.09 & 0.7 \\
Speaker-SFT-RL-OBJ & 0.93 & 0.6\% & 4.55 $\pm$ 0.11 & 0.81 \\ 
\hline
\end{tabular}
\label{table:exp_rl}
\end{table}

% experiments and results

The data created by human raters are labour demanded especially for single speaker RL training. Thus, it is worthy to analyze the utility of pairs generated by using just objective metrics. According to the evaluation results from Table \ref{table:exp_rl}, applying RL training onto the weights in Domain SFT using data rated by objective metrics do not bring significant improvement. However, we see much larger improvement from the results of \textit{Speaker-SFT-RL-OBJ}, though the results do not demonstrate the same level of improvement as in \textit{Speaker-SFT-RL-SUBJ}. We further analyze those experimental result by conducting a consistency analysis over data rated by objective metrics, regarding the human rated ones as ground truth. We find there is 64\% overlap in Speaker RL experiments, while there is only 55\% overlap in Domain RL experiments, which is consistent with the experimental results in Table \ref{table:exp_rl} for \textit{Domain-SFT-RL-OBJ} and \textit{Speaker-SFT-RL-OBJ}.

\subsubsection{Emotion Control Based on Instructions}
\par\noindent\textbf{Data preparation} 
Regarding the textual description, our professional data expert proposes three dimensions for annotating a voice recording: speaking emotion, speaking state, and speaking rhythm. Considering the difficulty and accuracy of annotation, the dimension of emotion is more distinctive compared to the other two dimensions, with the control of the remaining dimensions acting as supplementary control for the emotional dimension. We have established nine commonly recognized emotional directions(see in Table\ref{table:exp_emo}) and then described them using various synonymous natural language. We have annotated approximately 100 hours of audio data, with each audio clip's corresponding textual annotation cross-validated by three different experienced data annotators. All these annotated data are utilized for supervised fine-tuning on a large language model. 

% \begin{table}[htbp]
% % \caption{Comparison between Takin-TTS and Takin-TTS-EMO.}
% \caption{Examples of natural language control}
% \centering
% \begin{tabular}{C{100mm}cccc}
% \hline
% Speaking Emotion

% \hline
% Speaking State
% \hline
% Speaking
% \hline

% \end{tabular}
% \label{table:exp_emo}
% \end{table}

\par\noindent\textbf{Evaluation Metrics} 
The performance of TakinTTS-Instruct compared with TakinTTS-base is shown in table\ref{table:exp_emo} and table\ref{table:exp_takin_emo2}. To evaluate the accuracy of instruction controllability over speech synthesis, we randomly selected 50 different sentences with a fixed speaker prompt for each emotion, attempting to test whether different instructions could achieve the goal of controlling the emotional style of the synthesized audio. These sample instructions were derived from user inputs, such as: \textbf{"The speaker says it with a smile, in a tone that is somewhat familiar and at a fast speed, expressing a pleasant emotion."}  

%TODO: 展示情感控制的准确度
%1.情感控制准确度不差于takin-tts-base，接近或者超越的水平（MOS）
%2.更多维度的控制，解耦合之后的银色相似度不低于takin-tts-base
%TODO: MOS不低于原始model， 

% \begin{table}[htbp]
% % \caption{Comparison between Takin-TTS and Takin-TTS-EMO.}
% \caption{Comparison of emotion control accuracy ($\uparrow$) between Takin-TTS and Takin-TTS-EMO.}
% \centering
% \begin{tabular}{C{18mm}lllllllll}
% \hline
% System & Admire & Alert & Anger & Fear & Disgust & Joy & Sad & Surprise \\
% \hline
% TakinTTS-EMO & 0.45$\pm$0.03 &0.35$\pm$0.04 & 0.73$\pm$0.08 & 0.79$\pm$0.02 & 0.59$\pm$0.02 & 1.00$\pm$0.00 & 0.85$\pm$0.03 & 0.43$\pm$0.02 \\
% TakinTTS & 0.34$\pm$0.05 & 0.27$\pm$0.06 & 0.67$\pm$0.03 & 0.55$\pm$0.04 & 0.53$\pm$0.06 & 1.00$\pm$0.00 & 0.71$\pm$0.11 & 0.25$\pm$0.01 \\
% \hline
% \end{tabular}
% \label{table:exp_emo}
% \end{table}

\begin{table}[ht]
    \caption{Comparison of emotion control accuracy ($\uparrow$) between Takin-TTS and Takin-TTS-Instruct.}
    \centering
    \small
    \begin{tabular}{C{18mm}cc|C{18mm}cc}
        \hline
        Emotion & TakinTTS & TakinTTS-Instruct & Emotion & TakinTTS & TakinTTS-Instruct \\
        \hline
        Admire & 0.34$\pm$0.05 & 0.45$\pm$0.03 & Disgust & 0.53$\pm$0.06 & 0.59$\pm$0.02 \\
        Alert & 0.27$\pm$0.06 & 0.35$\pm$0.04 & Joy & 1.00$\pm$0.00 & 1.00$\pm$0.00 \\
        Anger & 0.67$\pm$0.03 & 0.73$\pm$0.08 & Sad & 0.71$\pm$0.11 & 0.85$\pm$0.03 \\
        Fear & 0.55$\pm$0.04 & 0.79$\pm$0.02 & Surprise & 0.25$\pm$0.01 & 0.43$\pm$0.02 \\
        \hline
    \end{tabular}
    \label{table:exp_emo}
\end{table}

The research results in Table \ref{table:exp_emo} indicate that TakinTTS-Instruct demonstrates strong controllability under various command inputs. Compared to the model before instruction finetuning, there is a significant improvement in the emotion similarity between the prompt and generated speech.

\begin{table}[htbp]

\caption{Indicator of Instruction control accuracy ($\uparrow$) between TakinTTS and TakinTTS-Instruct.}
\centering
\begin{tabular}{C{28mm}ccc}
\hline
System  & PER($\downarrow$)      
& MOS($\uparrow$) 
& SIM($\uparrow$)                       \\
\hline
TakinTTS-Instruct  & 1.9    & 4.48$\pm$0.13    & 0.78 \\
TakinTTS & 1.82   & 4.46$\pm$0.07    & 0.79  \\
\hline
\end{tabular}
\label{table:exp_takin_emo2}
\end{table}

Furthermore, the objective metrics in Table\ref{table:exp_takin_emo2} displays that the quality of the generated speechs from TakinTTS-Instruct system are no less than those of the benchmark Takin TTS, and even slightly outperformed.

\subsubsection{Efficient Inference and Serving} 

% 和来朋对齐 @yuguang

To generate speech with superior quality, we use auto-regressive LLMs and diffusion models in Takin, which are difficult and expensive to deploy. So we use various techniques and tricks to build an inference service with low latency and high concurrency. Our efforts on TTS task are as follows:
Since most of the computation is spent on LLM model inference, we deploy a separate service for LLM to maximize GPU utilization and throughput for token prediction. Flash attention\cite{flashattention,flashattention2} and paged attention\cite{pagedattention} techniques are used in the prefill and the decode phases respectively, to reduce the consumption of memory and computation. Mixed precision and quantization techniques such as GPTQ\cite{gptq} and AWQ\cite{awq} are also used to achieve further speedup. Besides, we adopt a suite of kernel-level optimizations, which leverage hardware-specific features and software techniques to accelerate critical computation kernels.
As described above, CSFT strategy is used to improve the stability of synthesis. But it is not practical to deploy separate inference services for different domains and speakers. So we support multiple LoRAs in the same service, as well as batch inference for different LoRAs. Streaming inference is applied to scenarios such as real-time interaction, and the first packet delay is less than 300 ms.

\subsection{Takin VC Experiments}
\subsubsection{Takin VC Datasets}

\textbf{Training Dataset} used in Takin VC training heavily overlaps with the data in the TTS dataset, including approximately 500,000 hours of web-scraped and internal data.

\textbf{Test Dataset}
We random select 100 out-of-set speaker speech data from the Internet. In addition, these speakers include different attributes such as gender, age, language, and emotion. Each speaker has about 1 to 3 sentences for different attributes. 
\subsubsection{Takin VC Performence}

\begin{table}[htbp]
% \caption{Comparison between Takin-TTS and Takin-TTS-EMO.}
\caption{Comparison of Takin-VC and baselines.}
\centering
\begin{tabular}{C{20mm}cccc}
\hline
System  & PMOS & SMOS & UTMOS & SIM  \\
\hline
DiffVC  & 3.34 $\pm$ 0.07    & 3.45 $\pm$ 0.07   & 3.48     & 0.61 \\
ValleVC $\diamond$  & 3.48 $\pm$ 0.05   & 3.53 $\pm$ 0.08   & 3.59     & 0.67 \\
NS2VC  & 3.31  $\pm$ 0.06  & 3.52  $\pm$ 0.07  & 3.45     & 0.54 \\
DDDMVC  & 3.56  $\pm$ 0.07  & 3.61 $\pm$ 0.07   & 3.67     & 0.69 \\
TakinVC & \textbf{4.02  $\pm$ 0.04}  & \textbf{4.07 $\pm$ 0.05}   & \textbf{4.16}    & \textbf{0.80}  \\
\hline
\end{tabular}
\begin{tablenotes}
    \centering
    \item $\diamond$ stands for utilizing VC models derived from open-source repositories.
\end{tablenotes}
\label{table:exp_takin_vc}
\end{table}

As shown in Table \ref{table:exp_takin_vc}, our proposed Takin VC scheme surpasses the baseline solution in terms of both sound quality and speaker similarity. Our experiments were conducted under conditions of large datasets to ensure the scheme's effectiveness on a large scale.

\subsection{Takin Morphing Experiments}
\par\noindent\textbf{Experimental Datasets} We trained Takin Morphing on a substantial corpus consisting of 20,000 hours of multilingual speech recordings in English and Chinese, consisting of in-house dataset alongside filtered portions of the WenetSpeech \cite{zhang2022wenetspeech} and LibriLight \cite{kahn2020libri}. 
To assess the performance of the proposed approach, we perform zero-shot speech synthesis and prosody transfer evaluations using in-house test sets which will be detailed below. 

\par\noindent\textbf{Evaluation Metrics} To conduct an in-depth analysis of the proposed Takin Morphing approach, various objective and subjective metrics are employed. To elaborate, PER and SIM are used as objective measures as well, while quality mean option score (QMOS) is employed to assess quality, clarity, naturalness, and high-frequency details, and similarity mean option score (SMOS) is used to measure speaker similarity with respect to timbre reconstruction and prosodic patterns for subjective evaluation.

\subsubsection{Zero-shot Speech Synthesis}

To examine the zero-shot speech synthesis performance of the proposed Takin Morphing, we first designed two distinct test sets, referred to as the objective and the subjective test sets. The objective test set includes 2,000 samples each from in-house English (EN) and Mandarin (ZH) speech corpora, while the latter comprises 200 highly expressive in-house samples in both EN and ZH as well. Notably, each sample in the subjective test set includes a reference utterance and a target utterance spoken by the same speaker. During inference, the Takin Morphing System generates speech for the target text using the reference speech as an audio prompt. The results are presented in Table \ref{table:takinmorphing1}.

\begin{table}[h]
\caption{Zero-shot speech synthesis results of Takin Morphing against real human speech.}
\centering

\begin{tabular}{cccccc}
\hline
Models       &  Language       & PER            &   SIM            & QMOS             & SMOS      \\ \hline
GT               &     EN      &  2.52\%        &   0.834     &    4.41 $\pm$ 0.08    &   4.28 $\pm$ 0.12     \\
Takin Morphing   &     EN      &  3.14\%        &   0.846     &    4.09 $\pm$ 0.07    &   4.04 $\pm$ 0.06     \\
\hline
GT               &     CN      &  2.16\%        &   0.879     &    4.43 $\pm$ 0.11    &   4.32 $\pm$ 0.09     \\
Takin Morphing   &     CN      &  3.05\%        &   0.884     &    4.13 $\pm$ 0.09    &   4.09 $\pm$ 0.08     \\
\hline
\end{tabular}
\label{table:takinmorphing1}
\end{table}

As shown in the table, on both Chinese and English test sets, our proposed Takin Morphing system achieved a performance level comparable to that of humans in terms of speech naturalness and speaker similarity. Notably, it slightly underperformed in the PER, QMOS, and SMOS metrics while achieving a higher score in the SIM metric. This outcome may be attributed to the fact that, even when both the real and reference speech originate from the same speaker, variations in speaking style, background environment, and other factors may still exist. In this context, Takin Morphing, when generating the target speech, accurately captures the fine-grained characteristics of the reference speech through more sophisticated and advanced timbre modeling, thereby enabling a more consistent and precise reproduction of the reference speech.

\subsubsection{Prosody Transfer}

To validate the prosody transfer performance of Takin Morphing, we transfer the styles from our internal dataset to audio samples from our main platform. Specifically, we randomly select 20 speakers from the main platform and choose 50 sentences for each of them. Subsequently, for each sentence of the selected speakers, we randomly choose an emotional speech clip from the internal emotional dataset and use it as the prosodic reference. 

\begin{table}[h]
\caption{Prosody transfer performance of Takin Morphing against real human speech.}
\centering

\begin{tabular}{cccccc}
\hline
Models       &  Language       & PER            &   SIM            & QMOS             & SMOS      \\ \hline
GT               &     EN      &  4.96\%        &   0.823     &    4.19 $\pm$ 0.12    &   4.21 $\pm$ 0.09     \\
Takin Morphing   &     EN      &  5.32\%        &   0.835     &    3.94 $\pm$ 0.09    &   3.90 $\pm$ 0.07     \\
\hline
GT               &     CN      &  2.99\%        &   0.853     &    4.21 $\pm$ 0.09    &   4.24 $\pm$ 0.11     \\
Takin Morphing   &     CN      &  3.05\%        &   0.875     &    3.99 $\pm$ 0.06    &   3.92 $\pm$ 0.08     \\
\hline
\end{tabular}
\label{table:takinmorphing2}
\end{table}

Table \ref{table:takinmorphing2} presents all results for English (EN) and Chinese (CN), respectively.
% On the EN test set, we can observe that the Tarkin Morphing achieved human-level performance with similar content recognition accuracy, QMOS, SMOS, and better SIM score.
In terms of the objective evaluation, we can observe that Tarkin Morphing consistently achieved human-level performance with similar content recognition accuracy and better SIM score, highlighting the effectiveness of systematical design of our proposed approach. In subjective tests, Takin Morphing demonstrated a performance level in both English and Chinese that closely matches real human speech, with QMOS and SMOS scores both exceeding 3.9, underscoring the effectiveness of the proposed method in prosody interpolation. 
% More samples can be found at [URL].

% \section{Future Work}

\section{Applications}

\subsection{Audiobook Generation} % TODO: @jianhao 

Takin TTS shows a large superiority comparing to conventional neural speech synthesis methods \cite{Wavnet,deepvoice,tacotron2,fastspeech2,vits}, which revolutionizes the field of AI audiobook generation. Two distinct approaches to creating immersive audio experiences using Takin TTS are explored. In the first approach, purely AI-generated audio content is produced, where different AI-powered voices act as various characters, bringing the story to life with diverse and nuanced performances. This approach allows for a consistent and scalable production process, potentially reducing costs and time associated with traditional audiobook recording. The another approach combines AI and human voices, with Takin TTS handling narration while human voice actors take on the dialogue parts. This hybrid approach leverages the efficiency and consistency of AI-generated speech for descriptive passages while preserving the emotional depth and authenticity that human actors bring to character interactions. The AI-generated audiobook samples can be listened in our demo page.

\subsection{Voice Clone}
In recent years, zero-shot timbre cloning technology has achieved significant advancements in voice cloning and speech synthesis, and is widely used in various fields. In voice assistants and customer service robots, it provides a more natural interaction experience; in the fields of film and entertainment content production, it is used for dubbing and creating voices for animated characters; in voice memos and recordings, it clones the voices of celebrities for future preservation. In music production, it can mimic the timbre of specific instruments; in education and training, it creates learning materials with standard pronunciations; in medical and rehabilitation, it helps patients who have lost the ability to speak regain their voices. Additionally, historical reconstructions and museum exhibits benefit from this technology. Using Takin VC's technology, the model requires only a few seconds to tens of seconds of audio samples to generate high-quality simulated voices, greatly reducing the technical threshold and making the aforementioned applications possible.

\subsection{Talking head}
\vspace{-1mm}
\begin{figure}[h]
    \centering
    \includegraphics[height=3.0cm]{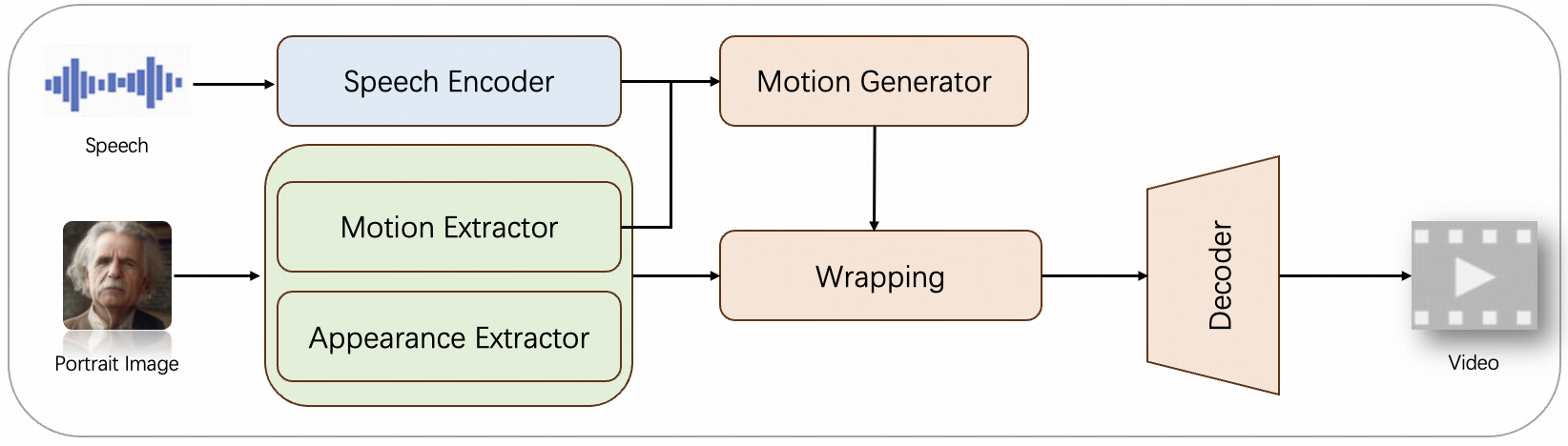}
    \caption{Talking Head Framework}
    \label{fig:ADA}
\end{figure}

By integrating LLM-based TTS systems with Portrait Animation technology, we can effortlessly create an interactive talking head system. The TTS system converts text to speech, while the Portrait Animation system generates expressive and temporally coherent animations synchronized with the speech, resulting in a lifelike animated character that communicates naturally and engagingly. The inference pipeline  is illustrated in Figure \ref{fig:ADA}.

\section{Authors (alphabetical order of family name) }
% NOTE: this author list is not complete version, needs to be refined

\begin{multicols}{3}
\begin{itemize}[noitemsep, leftmargin=*]
    \item Sijing Chen
    \item Yuan Feng
    \item Laipeng He
    \item Tianwei He
    \item Wendi He
    \item Yanni Hu
    \item Bin Lin
    \item Yiting Lin
    \item Yu Pan
    \item Pengfei Tan
    \item Chengwei Tian
    \item Chen Wang
    \item Zhicheng Wang
    \item Ruoye Xie
    \item Jixun Yao
    \item Quanlei Yan
    \item Yuguang Yang
    \item Jianhao Ye
    \item Jingjing Yin
    \item Yanzhen Yu
    \item Huimin Zhang
    \item Xiang Zhang
    \item Guangcheng Zhao
    \item Hongbin Zhou
    \item Pengpeng Zou
\end{itemize}
\end{multicols}

% \section*{References}

% \bibliographystyle{unsrtnat}
\bibliographystyle{unsrt}
\bibliography{refs}

\end{document}